\documentclass[12pt,letterpaper]{article}
\title{Two-stage differences in differences}
\author{John Gardner\thanks{Department of Economics, University of Mississippi, University, MS. {\tt jrgardne@olemiss.edu}. I thank Kyle Butts, Tao Wang, and participants at the Western Economics Association International conference for helpful comments and suggestions.}}
\date{This version: April 2021}
\usepackage{amsthm,amsmath,amsfonts,amssymb}
\usepackage{graphicx}
\usepackage{natbib}
\usepackage[margin=1in]{geometry}
\usepackage{setspace}
\begin{document}
%%%%%%%%%%%%%%%%%%%%%%%%%%%%%%%%%%%%%%%%%%%%%%%%%%%%%%%%%%%%%%%%%%%%%%%%%%%%%%%%
\maketitle

\thispagestyle{empty}
\setcounter{page}{0}

\begin{abstract}
\onehalfspacing
A recent literature has shown that when adoption of a treatment is staggered and average treatment effects vary across groups and over time, difference-in-differences regression does not identify an easily interpretable measure of the typical effect of the treatment. In this paper, I extend this literature in two ways. First, I provide some simple underlying intuition for why difference-in-differences regression does not identify a group$\times$period average treatment effect. Second, I propose an alternative two-stage estimation framework, motivated by this intuition. In this framework, group and period effects are identified in a first stage from the sample of untreated observations, and average treatment effects are identified in a second stage by comparing treated and untreated outcomes, after removing these group and period effects. The two-stage approach is robust to treatment-effect heterogeneity under staggered adoption, and can be used to identify a host of different average treatment effect measures. It is also simple, intuitive, and easy to implement. I establish the theoretical properties of the two-stage approach and demonstrate its effectiveness and applicability using Monte-Carlo evidence and an example from the literature. 


\end{abstract}

\clearpage

\section{Introduction}

The difference-in-differences methodology has become an essential part of the applied empiricist's toolkit for evaluating treatment effects. Recently, however, an insightful literature has shown that, when the adoption of the treatment by different groups is staggered over time, and the average effects of the treatment vary over groups and across time, the usual difference-in-differences regression specification does not identify a readily interpretable measure of the typical effect of the treatment  \citep[see][]{borusyak2017, athey2018, goodman-bacon2018, dechaisemartin2020, imai2020, sun2020}. Given the popularity, and the utility, of differences in differences, this is disconcerting.

In this paper, I extend the literature on difference-in-differences with staggered adoption and heterogeneous treatment effects in two ways. First, I present some clarifying intuition for why differences-in-differences may not identify the average effect of the treatment on the treated. Motivated by this intuition, I then develop a simple two-stage alternative to difference-in-differences regression that is robust to treatment-effect heterogeneity when adoption is staggered.

Presumably, part of why difference-in-differences regression is ubiquitous in settings with multiple groups and time periods is because it seems natural that it should identify the average effect of the treatment on the treated.\footnote{As \citet{borusyak2017} write, ``...there is a perception that [differences in differences] should estimate average treatment effects with some reasonable weights.''} Since it \emph{does} identify the average of heterogeneous treatment effects as long as those effects are distributed identically across treatment groups and periods (a condition that holds automatically in the classical two-group, two-period setting), this is an understandable misconception. When those distributions are not identical, however, conditional mean outcomes are not linear in group, period, and treatment status, so the standard differences-in-differences regression model is misspecified, and therefore does not identify the average effect of the treatment on the treated.

The helps explain why differences-in-differences may not identify the average effect of the treatment on the treated, but says little about what it does identify. Several papers have provided alternative representations of the difference-in-difference regression estimand. \cite{borusyak2017} show that regression difference-in-differences identifies a regression-weighted mean of the average effect of the treatment in each post-treatment period, and \cite{dechaisemartin2020} show that all two-way fixed-effects regression estimates (which include difference-in-differences regressions as a special case) identify weighted averages of group- and period-specific average treatment effects. Since the weights in both of these representations can be negative, the difference-in-differences estimand may be difficult to interpret. \citet{goodman-bacon2018} further shows that the regression difference-in-differences estimate represents a weighted average of all two-group, two-period differences in differences, which under parallel trends identifies a combination of weighted averages of group$\times$period-specific average treatment effects and changes over time in those effects. As I discuss below, these decomposition results can be interpreted as describing how misspecified difference-in-differences regression models project heterogeneous treatment effects onto group and period fixed effects, rather than treatment status itself.

There are several alternatives to the difference-in-differences regression approach that are robust to heterogeneity across groups and periods when treatment adoption is staggered. Following \cite{gibbons2017}, \cite{borusyak2017}, \cite{callaway2018}, and \cite{sun2020}, one alternative is to estimate separate average treatment effects for each group and period, which can then be aggregated to form measures of the overall effect of the treatment (\citealp{gibbons2017}, suggest an approach like this for fixed effects models, \citealp{borusyak2017}, suggest such a solution for difference-in-differences models in which the duration-specific effects of the treatment are identical across groups, as do \citealp{callaway2018}, for the case when treatment effects vary by group and duration, and \citealp{sun2020}, in the event-study context).\footnote{One advantage of these approaches is that they enable researchers to estimate a variety of different summary measures of treatment effects. The method developed by \citet[][cf.\ \citealp{abadie2005}]{callaway2018} also accommodates covariates (which I mostly abstract away from) more flexibly than traditional regression-based methods.} Another is the ``stacked'' difference-in-differences approach \cite[see, e.g.,][]{gormley2011, deshpande2019, cengiz2019}, which attempts to transform the staggered adoption setting to a two-group, two-period design (in which difference in differences identifies the overall average effect of the treatment on the treated) by stacking separate datasets containing observations on treated and control units for each treatment group.\footnote{In Appendix A, I show that this estimator identifies an average of group-specific average treatment effects, weighted by the relative sizes of the group-specific datasets and the variance of treatment status within those datasets.} 

I develop an alternative, two-stage regression approach to identification that is robust to treatment-effect heterogeneity when adoption of the treatment is staggered. In its simplest form, the first stage of the procedure consists of a regression of outcomes on group and period fixed effects, estimated using the subsample of untreated observations. In the second stage, the estimated group and period effects are subtracted from observed outcomes, and these adjusted outcomes are regressed on treatment status. Under the usual parallel trends assumption, this procedure identifies the overall average effect of the treatment on the treated (i.e., across groups and periods), even when average treatment effects are heterogeneous over groups and periods. 

The two-stage approach can be adapted to recover a variety of treatment effect measures, and extended to event-study analyses of pre-trends and duration-specific average treatment effects. In the event-study context, this same approach has also been suggested by \citet{thakral2020} in their study of the how the anticipation of future consumption influences spending decisions. The two-stage estimators can be implemented easily, along with valid asymptotic standard errors, using standard statistical software, and with little programming beyond that required to estimate a regression. It is also simple, and preserves the intuition behind identification in the two-group, two-period case: it recovers the average difference in outcomes between treated and untreated units, after removing group and period effects. 

I motivate the problem with the difference-in-differences regression approach, and discuss the difference-in-differences regression estimand, in Section \ref{sec:motivation}. I introduce the two-stage approach and its properties in Section \ref{sec:2sdd}. I illustrate the performance of the two-stage approach using Monte Carlo evidence in Section \ref{sec:sim} and its application in Section \ref{sec:app}. I conclude in Section \ref{sec:conclusion}.

\section{Motivation}\label{sec:motivation}

\subsection{The problem with difference-in-differences regression}\label{sec:problems}

Difference-in-differences research designs attempt to identify the causal effects of treatments under the parallel or common trends assumption, which asserts that, absent the treatment, treated units would experience the same change in outcomes as untreated units. Mathematically, this amounts to the assumption that average untreated potential outcomes decompose into additive group and period effects. More formally, let $i$ index units (e.g., states or, with microdata, individuals) and $t$ index calendar time (often years). Further divide individuals and time into treatment groups $g \in \{0, 1, \dots, G\}$ and periods $p \in \{0, 1, \dots, P\}$ defined by the adoption of the treatment among successive groups, so that members of group 0 are untreated in all periods, only members of group 1 are treated in period 1, members of groups 1 and 2 are treated in period 2, and so on. Let $Y_{gpit}$, $Y_{1gpit}$ and $Y_{0gpit}$ denote the observed, treated, and untreated potential outcomes for the $i$th member of group $g$ during time $t$ of period $p$, $D_{gp}$ be an indicator for whether members of group $g$ are treated in period $p$, and $\beta_{gp}=E(Y_{1gpit}-Y_{0gpit}|g,p)$ denote the average causal effect of the treatment for members of $g$ in $p$. Under parallel trends, mean outcomes satisfy
\begin{equation}\label{eq:pt}
  E(Y_{gpit}|g, p, D_{gp}) = \lambda_g + \gamma_p + \beta_{gp} D_{gp}.
\end{equation}

The idea behind differences in differences is to eliminate the permanent group effects $\lambda_g$ and secular period effects $\gamma_p$ in order to identify the average effect of the treatment. In the classic setup, there are only two periods (pre and post) and two groups (treatment and control). In this setting, within-group differences over time eliminate the group effects and within-period differences between groups eliminate the period effects. Hence the between-group difference in post-pre differences (i.e., the difference in differences) identifies the average effect of the treatment for members of the treatment group during the post-treatment period.

The two-period, two-group difference-in-differences estimate can be obtained manually, by calculating each of the four group$\times$period averages and taking differences, or via a regression of outcomes on group and period fixed effects and a treatment-status indicator:
\begin{equation}\label{eq:dd}
  Y_{gpit} = \lambda_g + \gamma_p + \beta D_{gp} + \varepsilon_{gpit}.
\end{equation}
It follows from \eqref{eq:pt} that the coefficient on $D_{gp}$ in \eqref{eq:dd} identifies the average effect of the treatment on the treated, or $\beta_{11}=E(Y_{1gpit}-Y_{0gpit}|D_{gp}=1)$.\footnote{There are several equivalent variations on this regression. Specification \eqref{eq:dd} is identical to a regression of outcomes on an indicator $Post_{it}$ for whether $t$ occurs in the post-treatment period, an indicator $Treat_{it}$ for whether $i$ belongs to the treatment group, and an interaction between the two. Often, the group and period effects $\lambda_g$ and $\gamma_p$ in \eqref{eq:dd} are replaced with individual and time effects $\lambda_i$ and $\gamma_t$. By the Frisch-Waugh-Lovell theorem, the coefficient on $D_{gp}$ in \eqref{eq:dd} can be obtained by regressing $Y_{gpit}$ on the residuals from a regression of treatment status on group and period effects. Since treatment status only varies by group and period, these residuals are the same as those from a regression of treatment status on individual and time effects, so the coefficients on treatment status from both specifications are identical (despite the fact that the latter model is misspecified for $E(Y_{it}|i,t,D_{it})$).}

The regression approach suggests a natural way to extend the differences-in-differences idea to settings with multiple groups and time periods. Unfortunately, as several authors have noted \citep{borusyak2017, goodman-bacon2018, athey2018, dechaisemartin2020, imai2020}, when the average effect of the treatment varies across groups and over periods, the coefficient on $D_{gp}$ in specification \eqref{eq:dd} does not always identify an easily interpretable measure of the ``typical'' effect of the treatment. Although this result is now well established, because it is also somewhat counterintuitive, it bears further clarification.

While there are multiple ways to think about the typical effect of the treatment when that effect varies across groups and over time (see Section \ref{sec:estimands} below), an obvious candidate is the average $E(\beta_{gp} | D_{gp}=1)=E(Y_{1gpit}-Y_{0gpit}| D_{gp}=1)$ of group- and period-specific average treatment effects, taken over all units that receive the treatment and all times during which they receive it. This is what differences in differences identifies in the two period, two group case, and is probably what most people have in mind when they think about heterogenous treatment effects averaged over groups and time. Using this measure, parallel trends can be expressed as
\[
  E(Y_{gpit} | g, p, D_{gp}) = \lambda_g + \gamma_p + E(\beta_{gp}| D_{gp}=1) D_{gp}
  + [\beta_{gp} - E(\beta_{gp}| D_{gp}=1)] D_{gp}.
\]
The difficulty with the regression approach arises because, except in special cases, the ``error term'' $[\beta_{gp} - E(\beta_{gp} | D_{gp}=1)] D_{gp}$ in this expression varies at the group$\times$period level, and is not mean zero conditional on group membership, period, and treatment status. Consequently, the conditional expectation $E(Y_{gpit} | g, p, D_{gp})$ is not, in general, a linear function of those variables (at least, not one in which the coefficient on $D_{gp}$ is $E(\beta_{gp} | D_{gp}=1)$), and \eqref{eq:dd} is misspecified.\footnote{Cf.\ \citet[][Theorem 3.1.4]{angrist2009}: ``Suppose the CEF is linear. Then the population regression function is it.''} In contrast to the two-group, two-period case, the coefficient on $D_{gp}$ from the regression differences-in-differences specification \eqref{eq:dd} does not identify the average difference $E(\beta_{gp} | D_{gp}=1)$ in outcomes between treated and untreated units after removing group and period fixed effects, unless those average effects are independent of group and period (in which case $\beta_{gp}=E(\beta_{gp} | D_{gp}=1)=\beta$, which the coefficient on $D_{gp}$ recovers). Outside of this special case, when average treatment effects vary across groups and periods, and the adoption of the treatment by different groups is staggered over time, difference-in-differences regression does not recover a simple group$\times$period average treatment effect.

\subsection{The difference-in-differences regression estimand}\label{sec:dd-estimand}

The preceding argument clarifies why difference-in-differences regression may not recover a readily interpretable measure of the average effect of the treatment, but says nothing about what it does identify. An easy answer to that question is that \eqref{eq:dd} identifies the linear projection of average outcomes onto group and period effects and a treatment indicator (which differs $E(Y_{gpit}|g, p, D_{gp})$ when that function is nonlinear). To provide additional insight into the difference-in-differences estimand, it can be shown that, under parallel trends, the coefficient on $D_{gp}$ from the difference-in-differences regression specification \eqref{eq:dd} identifies
\[
  \beta^* = \sum_{g=1}^G \sum_{p=g}^{P} \omega_{gp} \beta_{gp},
\]
where
\begin{equation}\label{eq:omega}
  \omega_{gp} = \frac{\{[1 - P(D_{gp}=1 | g)] - [P(D_{gp}=1|p)-P(D_{gp}=1)]\}P(g, p)}
  {\sum_{g=1}^G \sum_{p=g}^{P} \{[1 - P(D_{gp}=1 | g)] - [P(D_{gp}=1|p)-P(D_{gp}=1)]\}P(g, p)},
\end{equation}
$P(D_{gp}=1|p)$ is the fraction of units that are treated in period $p$, $P(D_{gp
}=1|g)$ is the fraction of periods in which members of group $g$ are treated, $P(D_{gp}=1)$ is the fraction of unit$\times$times that are treated, and $P(p, g)$ is the population share of observations that correspond to group $g$ and period $p$. This representation can be obtained from Theorem 1 of \citet{dechaisemartin2020}, although I present an alternative derivation based on population regression algebra in Appendix A.\footnote{An immediate implication of \eqref{eq:omega} is that the weights must sum to one. Another is that $\omega_{11}=1$ when there is only one treatment group, so the regression differences-in-differences specification \eqref{eq:dd} identifies the average effect of the treatment on the treated, as noted above.}

Appearances notwithstanding, this weighting scheme is deeply intuitive. Specification \eqref{eq:dd} assumes a conditional expectation function that is linear in group, period, and treatment status. When misspecified, it will attribute some of the heterogeneous impacts of the treatment to group and period fixed effects.\footnote{This is consistent with the intuition provided by \citet{borusyak2017}, \citet{goodman-bacon2018} and \citet{dechaisemartin2020} that \eqref{eq:dd} uses already-treated units as controls for newly treated ones.} The longer a group's observed treatment duration (i.e., the greater $P(D_{gp}=1 | g)$ is), the more that group's treatment effects will be absorbed by group fixed effects. Likewise, the greater the probability of being treated in a particular period (i.e., the greater $P(D_{gp}=1|p)$ is), the more treatment effects experienced during that period will be absorbed by period effects. Larger groups also receive more weight. I illustrate these phenomena by simulation in Section \ref{sec:sim} and empirically in Section \ref{sec:app}.

\section{A two-stage approach}\label{sec:2sdd}

In the two-period, two-group case, differences-in-differences regression recovers the difference in outcomes between treated and untreated units after removing group and period effects, which under parallel trends represents the average effect of the treatment on the treated. This is not true when there are multiple groups and periods, since in this case \eqref{eq:dd} is misspecified for conditional mean outcomes. However, this observation suggests a simple two-stage average treatment effect estimator for the multiple group and period case. As long there are untreated and treated observations for each group and period, $\lambda_g$ and $\gamma_p$ are identified from the subpopulation of untreated groups and periods. The overall group$\times$period average effect of the treatment on the treated is then identified from a comparison of mean outcomes between treated and untreated groups, after removing the group and period effects. 

Following this logic, a two-stage estimation procedure is
\begin{enumerate}

\item Estimate the model
  \[ Y_{gpit} = \lambda_g + \gamma_p + \varepsilon_{gpit} \]
on the sample of observations for which $D_{gp}=0$, retaining the estimated group and time effects $\hat{\lambda}_g$ and $\hat{\gamma}_p$.\footnote{There are variations on this stage of the procedure. It is not necessary to estimate the fixed effects using only untreated observations, any correctly specified conditional mean function will do. For example, one could also use a specification that includes interactions between treatment status and period indicators, or one with  group$\times$period-specific treatment-status indicators. Since these variations utilize the entire sample to estimate the group and period effects, they may be more efficient. Because treatment status only varies at the group$\times$period level, the group and period effects can also be replaced with fixed effects for individual units and time periods.}

\item Regress adjusted outcomes $Y_{gpit} - \hat{\lambda}_g - \hat{\gamma}_p$ on $D_{gp}$.
\end{enumerate}

Since parallel trends implies that
\[
  E(Y_{gpit}| g, p, D_{gp}) - \lambda_g - \gamma_p = \beta_{gp} D_{gp} 
  = E(\beta_{gp} | D_{gp}=1) D_{gp} + [\beta_{gp} - E(\beta_{gp} | D_{gp}=1)] D_{gp},
\]
where $E\{[\beta_{gp} - E(\beta_{gp} | D_{gp}=1)] D_{gp} | D_{gp}\}=0$, this procedure identifies $E(\beta_{gp} | D_{gp}=1)$, even when the adoption and average effects of the treatment are heterogenous with respect to groups and periods.\footnote{A simple way to allow for dependence on time-varying covariates under this approach is to include them in both regressions, which is analogous to the way regression difference-in-differences analyses typically incorporate covariates. However, as \citet{callaway2018} note, this implicitly assumes that the treatment does not influence the covariates and that the covariates do not influence the effect of the treatment. In principle, the two-stage approach can accommodate their more stringent notion of conditional parallel trends by including interactions between covariates (time-varying or not) in the first stage (i.e., via the specification $Y_{gpit}=\lambda_g + \gamma_p + X_{gpit}'\delta_{pt} + \varepsilon_{gpit}$) and either adjusting outcomes in the second stage according to the last pre-treatment realization $X^\ast_{gpit}$ of the covariates for unit $(g,i)$ as of time $(p,t)$ (i.e., using $Y_{gpit} - \lambda_g - \gamma_p - {X^\ast_{gpit}}'\delta_{pt}$) or by controlling for $X^\ast_{gpit}$ via conventional propensity-score reweighting methods (or a doubly robust combination of methods), though this remains more parametric than their inverse-probability-weighting approach.}

Although I prove consistency and unbiasedness in the Appendix, the intuition is straightforward. Unbiasedness of the first- (and hence second-) stage estimates follows from standard arguments. If the $(g,p,i,t)$ represent individual-level observations, so does the consistency of the first stage for the group and period effects. Otherwise, if they represent aggregates (e.g., state$\times$year-level sample averages), then as the number of individual-level observations used to calculate them increases, the first-stage is equivalent to a population regression of $(g,p,i,t)$-level means on group and period effects, and hence consistent for those effects. In either case, the consistency of the second-stage for $E(\beta_{gp}|D_{gp}=1)$ follows from the consistency of the first stage for the group and period effects.\footnote{Also note that restricting the sample to untreated observations does not introduce sample-selection bias because the selection is with respect treatment status, which is a deterministic function of the group and period variables included in the model.}

\subsection{Estimands}\label{sec:estimands}

Implemented as described, the two-stage difference-in-differences estimator identifies $E(\beta_{gp} | D_{gp}=1)$, where the expectation is implicitly taken with respect to all observed units and periods. This expectation can be expressed as
\begin{equation}\label{eq:gp-avg}
  E(\beta_{gp} | D_{gp}=1) = \sum_{g=1}^G \sum_{p=g}^{P} \beta_{gp} P(g,p| D_{gp}=1).
\end{equation}
While this is a natural summary measure of group$\times$period-specific average treatment effects, since it reflects the uneven progression of different groups through the course of the treatment, it may not be especially informative for program evaluation and policy analysis. For example, even if the effects of the treatment are identical across groups, this measure will put more weight on groups that are in early stages of the treatment.\footnote{When treatment effects vary by group, it is unclear whether any summary measure will be informative about how the treatment might affect future groups. External validity with this type of heterogeneity is inherently challenging.} \citet{callaway2018} provide a much richer discussion of how heterogeneous average treatment effects can be summarized.

If there is some treatment duration $\bar{P}$ which all, or a subset of, groups has completed, then an alternative summary measure is the $\bar{P}$-period average 
\begin{equation}\label{eq:pbar-avg}
	\sum_{g=1}^G \sum_{p=g}^{g+\bar{P}-1} \beta_{gp} P(g|D_g=1)/\bar{P},
\end{equation}
where $P(g| D_g=1)$ is the fraction of treated units that belong to group $g$. Because this measure averages the group-specific average effects of the treatment for a common set of completed durations, it may provide a more balanced picture of the typical effect of the treatment, although it ignores the effects of the treatment for durations longer than $\bar{P}$ periods. The two-stage procedure can be modified to identify this measure by restricting the sample used in the second step to untreated observations and treated observations with durations no greater than $\bar{P}$.

\subsection{Event studies}\label{sec:es}

Difference-in-differences analyses are often accompanied by event-study regressions of the form
\begin{equation}\label{eq:es}
  Y_{gpit} = \lambda_g + \gamma_p + \sum_{r=-R}^P \beta_r D_{rgp} + \varepsilon_{gpit},
\end{equation}
where for $r\le 0$ the $D_{rgp} \in \{D_{-Rgp},\dots,D_{0gp}\}$ are $(r+1)$-period leads of treatment adoption, and for $r > 0$ the $D_{rgp} \in \{D_{1gp},\dots,D_{Pgp}\}$ are indicators for $r$-period treatment durations.\footnote{In event-study regressions, it is common practice to use calendar times $t$ in place of more coarse treatment periods $p$.} In principle, such regressions serve a dual purpose. First, they can be used to show how the effect of the treatment evolves over the course of the treatment. Second, the coefficients on the treatment adoption leads can be used as placebo tests for the plausibility of parallel trends.

\citet{sun2020} show that, when duration-specific average treatment effects vary across groups¢, event-study regressions suffer from the same problem as difference-in-differences regressions. This can be seen using an argument similar to the one presented for difference-in-differences regressions in Section \ref{sec:problems}. Let $Y_{rgpit}$ denote potential outcomes after $r$ periods of treatment, and $\beta_{rgp} = E(Y_{rgpit}-Y_{0gpit} | g, p, D_{rgp}=1)$ be the average effect of being treated for $r$ periods for members of group $g$ in time period $p$.\footnote{There is a one-to-one correspondence between duration- and period-specific treatment effects. In terms of the group$\times$period average treatment effects $\beta_{gp}$, the duration-specific effects satisfy $\beta_{rgp}=\beta_{g,p-g+1}$. While in principle the duration-specific average treatment effects for each group might vary over time, in practice we only ever observe each treatment duration at most once for each group.} Under parallel trends, we can write
\[
  E[Y_{gpit} | g, p, (D_{rgp})] = \lambda_g + \gamma_p + \sum_{r=1}^{P} E(\beta_{rgp}|D_{rgp}=1) D_{rpg} + 
  \sum_{r=1}^{P} [\beta_{rgp} - E(\beta_{rgp}|D_{rgp}=1)]D_{rgp},
\]
where, in general, $E\{\sum_{r=1}^{P} [\beta_{rgp} - E(\beta_{rgp}|D_{rgp}=1)]D_{rgp} | g, p, (D_{rgp})\} \ne 0$. Hence, mean outcomes are not necessarily linear in group, period, and treatment-duration indicators, so the coefficients on the $D_{rgp}$ from \eqref{eq:es} do not identify the average effects of being treated for $r$ periods. \cite{sun2020} further show that the coefficients on the adoption leads and duration indicators identify weighted averages of all of the group$\times$period-specific average treatment effects. An important consequence of this is that the coefficients on the treatment-adoption leads $D_{rgp}$, $r \le 0$, may be nonzero even if trends are, in fact, parallel. 

The two-stage procedure developed above can be extended to the event-study setting by amending the second stage of the procedure to:
\begin{enumerate}
\item[2'.]  Regress $Y_{gpit} - \hat{\lambda}_g - \hat{\gamma}_p$ on $D_{-Rgp},\dots,D_{0gt},\dots,D_{Pgp}$.
\end{enumerate}
Following the logic of the previous section, because $E[Y_{gpit}|g, p, (D_{rgp})]
 - \lambda_g - \gamma_p$ is linear in the $D_{rgp}$, the coefficients on the $D_{rgp}$ identify the average effects $E(\beta_{rgp} | D_{rgp}=1)$.\footnote{This expectation is taken over all groups with durations of at least $r$. Since under staggered adoption the completed treatment duration varies by group, the groups over which these duration-specific effects are averaged will vary across durations. These averages are also what the interaction-weighted estimator proposed by \citet{sun2020} identifies. If all groups are treated for at least $\bar{P}$ periods, an alternative is to exclude observations corresponding to treatment durations longer than $\bar{P}$ periods from the second-stage sample, in which case the two-stage approach identifies duration-specific treatment effects, averaged over all groups.}

\subsection{Inference}

The standard errors for the two-stage estimators need to be adjusted to account for the fact that the dependent variable $Y_{gpit}-\hat{\lambda}_g - \hat{\gamma_p}$ in the second-stage is generated using estimates obtained from the first stage of the procedure \citep{dumont2005}. The asymptotic distribution of the second-stage estimates can obtained by interpreting the two-stage procedure as a joint GMM estimator \citep{hansen1982}.

Let $W_{gpit}=[Y_{gpit}, (1(g)_{gpit}), (1(p)_{gpit}), D_{gp}]$ denote the data for observation $(g,p,i,t)$, consisting of the outcome $Y_{gpit}$, the $G$-vector of group-membership indicators $(1(g)_{gpit})$, a $P$-vector $(1(p)_{gpit})$ of period indicators for periods $p\in\{1,\dots,P\}$, and the treatment-status indicator $D_{gp}$. Let $\lambda$ be the $G$-vector of group fixed effects, $\gamma$ the $P$-vector of period fixed effects, and $\beta$ the group$\times$period average treatment effect. The two-stage difference-in-differences estimator solves the population analog of the moment condition 
\[
  E[f(\lambda, \gamma, \beta; W_{gpit})] = 
  E\begin{bmatrix}
  	[Y_{gpit} - (1(g)_{gpit})'\lambda - (1(p)_{gpit})'\gamma][(1(g)_{gpit}),(1(p)_{gpit})]'(1-D_{gp}) \\
  	[Y_{gpit} - (1(g)_{gpit})'\lambda - (1(p)_{gpit})'\gamma - \beta D_{gp}] D_{gp}
   \end{bmatrix}
  =0.
 \]

By Theorem 6.1 of \citet[][cf. \citealp{newey1984}]{newey1994}, and under standard regularity conditions, $\sqrt{N} (\hat\beta - \beta) \overset{a}{\sim} N(0, v)$, where $v$ is the last element of
 \[
   E\left[\frac{\partial f(\lambda, \gamma, \beta; W_{gpit})}{\partial (\lambda, \gamma, \beta)}\right]^{-1}
   E[f(\lambda, \gamma, \beta; W_{gpit}) f(\lambda, \gamma, \beta; W_{gpit})']
   E\left[\frac{\partial f(\lambda, \gamma, \beta; W_{gpit})}{\partial (\lambda, \gamma, \beta)}\right]^{-1'}.
 \]
Asymptotics for variations on the two-stage difference-in-differences estimator (such as the event-study version) are similar.

The preceding expression can be used to manually correct the estimated second-stage variances for the use of a generated dependent variable. With modern statistical software, a simpler approach is to estimate both stages of the procedure simultaneously using a GMM routine. In Appendix B, I provide example Stata syntax that shows how to implement the two-stage difference-in-differences approach (with valid asymptotic standard errors) via GMM.

\section{Simulations}\label{sec:sim}

To illustrate the effectiveness of the two-stage approach, I conduct two Monte Carlo studies. For each study, I simulate 250 datasets, each consisting of observations on 50 units over 10 periods. Unit-level outcomes are determined by 
\[
  Y_{gpit} = \lambda_i + \gamma_t + \beta_{gp} D_{gp} + \varepsilon_{gpit},
\]
with $\lambda_i, \varepsilon_{it} \sim N(0,1)$. I assume that the average effect of the treatment varies by group and period, with the treatment effects for each group stabilizing by the fourth period.\footnote{The vectors of average treatment effects for the first four periods are (2, 4, 6, 8), (1, 2, 3, 4), and (.5, 1, 3, 3.5) for groups one, two, and three.} For the first study, three treatment groups, each consisting of five units, adopt the treatment at times 4, 5, and 6, respectively. The only difference for the second study is that the sizes of the treatment groups vary, consisting of 5, 15, and 10 units.

\begin{figure}

\caption{Simulated DD weights\label{fig:sim-wts}}

\medskip{}

\begin{centering}
\includegraphics{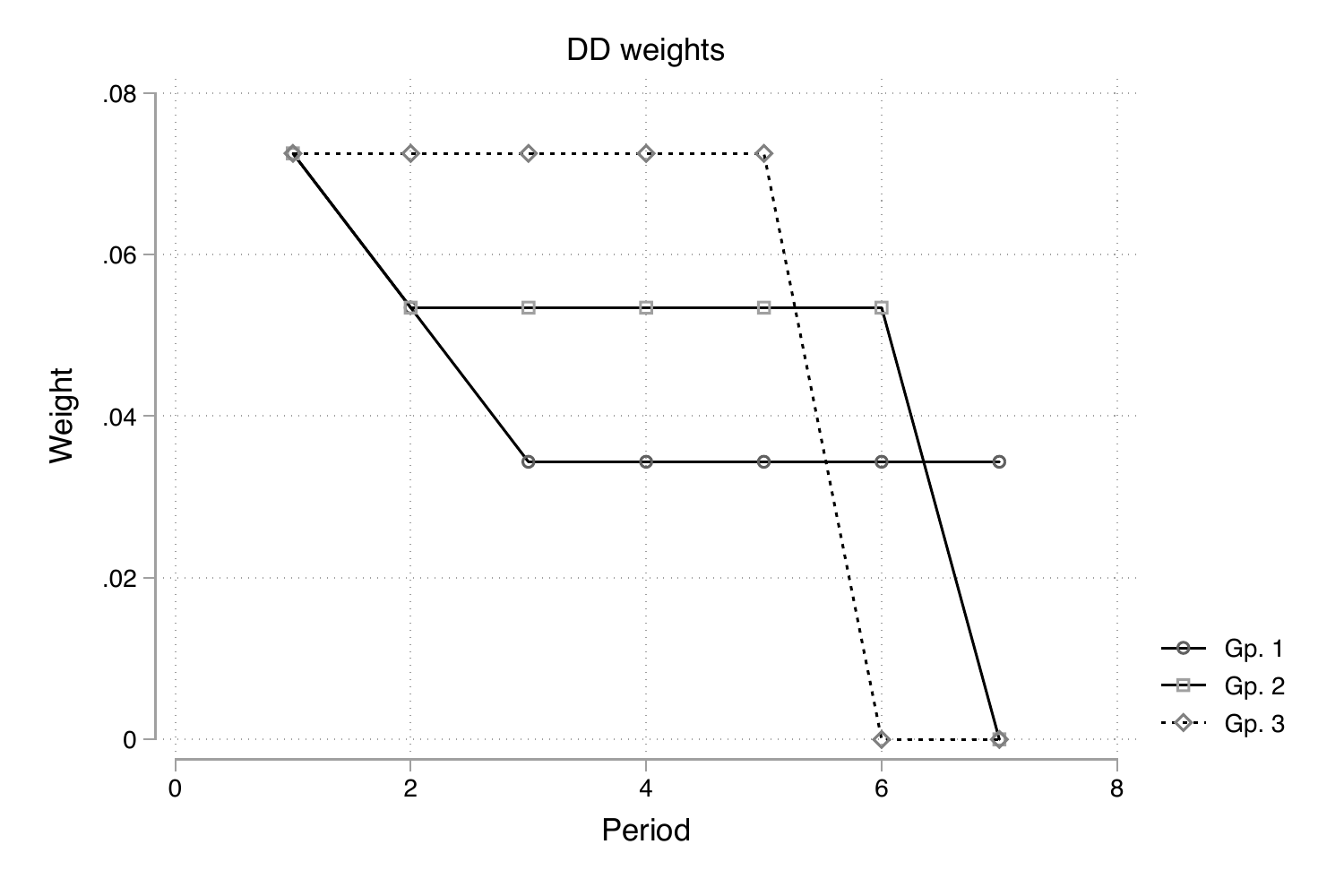}
\par\end{centering}
\end{figure}

Focusing initially on the first simulation, Figure \ref{fig:sim-wts} plots the weights that the regression difference-in-differences specification \eqref{eq:dd} places on each of the group$\times$period-specific treatment effects (here, I have aligned the weights by the calendar time, rather than the time since adoption). For the first two groups, the weight that \eqref{eq:dd} places on period-specific treatment effects decreases with each successive period, until the final group adopts the treatment and the treated share of units stabilizes. After this stabilization, treatment effects for earlier groups, who are treated for more periods, receive less weight. This is consistent with the theoretical predictions and intuition from Section \ref{sec:dd-estimand}.

Table \ref{tab:sim} presents the means and standard deviations of estimates of several average treatment effect measures. Results for the first simulation are presented in the first column. The first row of the top panel of the table presents the true group$\times$period average treatment effect (i.e., \eqref{eq:gp-avg}). The difference-in-differences regression estimate, which as \eqref{eq:omega} shows and Figure \ref{fig:sim-wts} illustrates, tends to put more weight on earlier treatment durations, understates the true average effect considerably. 

The third row summarizes estimates of the group$\times$period average, obtained by estimating models with separate treatment-status indicators for each group and period (as well as unit and time fixed effects), then aggregating the group$\times$period-specific average effects according to the empirical distribution of treated groups and periods (i.e., the sample analog of \eqref{eq:gp-avg}). Finally, the fourth row summarizes two-stage difference-in-difference regression estimates, obtained following the procedure outlined in Section \ref{sec:2sdd}. In this case, the two-stage estimates are indistinguishable from the aggregated estimates. Both perform well for the true group$\times$period average treatment effect.

As I note in Section \ref{sec:estimands}, the group$\times$period average treatment effect might not be the best way to summarize the effect of the treatment. Since all units in this simulation are treated for at least four periods, it is possible to estimate a four-period treatment effect, averaged across all treated groups. The first row of the bottom panel of Table \ref{tab:sim} presents the true four-period average.\footnote{Because the difference-in-difference regression estimate represents a weighted average of all observed group$\times$period treatment effects, it is not directly comparable to this four-period average.} The second row presents estimates aggregated from regressions that include group$\times$period-specific treatment indicators (i.e., the sample analog of \eqref{eq:pbar-avg}). 

The third row presents stacked difference-in-differences regression estimates of this average. To implement the stacked estimator, for each treated group I create a new dataset spanning two periods before and four periods after that group adopts the treatment, consisting of observations on the treatment group and the group of units that never receives the treatment. I then stack these group-specific datasets and regress outcomes on treatment status and dataset-specific group and period effects.\footnote{In some applications, treated units who have not yet adopted the treatment are also included as controls in each group-specific dataset. While it is possible to use the stacked approach to estimate a weighted group$\times$period average treatment effect, in most applications the group-specific datasets include the same number of treatment periods for each group.}

Finally, the fourth row presents two-stage difference-in-difference estimates, obtained by restricting the second stage to the sample of observations with treatment durations no greater than four periods. Here, the aggregated, stacked, and two-stage estimates are identical; all are centered closely on the true four-period average effect.

The sole difference between the first and second studies is that, in the second, the sizes of the treatment groups vary. The results for this study are presented in the second column of the table. Here, the relative performance of the aggregated and two-stage estimators for the group$\times$period average treatment effect is similar to the first study. For the four-period average, the aggregated and two-stage estimates are again identical, and close to the true average. In contrast, the stacked estimator, which weights each group's average treatment effect by dataset-specific treatment variance and sample size (see Appendix A), slightly overstates the true average. 

\begin{figure}
\caption{Simulated event studies\label{fig:sim-es}}

\medskip{}

\begin{centering}
\includegraphics[height=0.9\textheight]{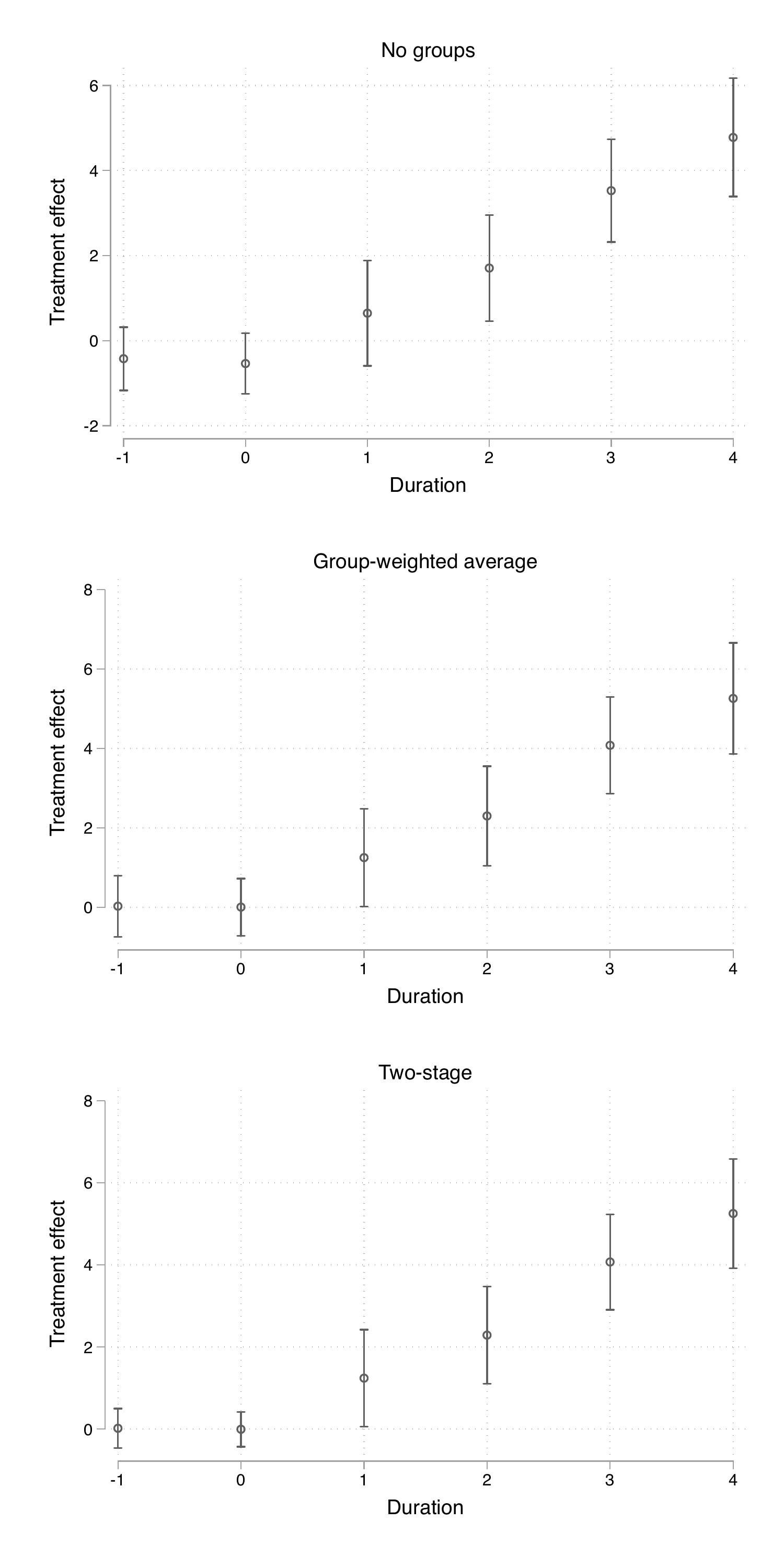}\medskip{}
\par\end{centering}
Notes: Means and $\pm$ 2$\times$standard deviations from 250 simulations.

\end{figure}

The top panel of Figure \ref{fig:sim-es} summarizes estimates from event-study regressions that include two leads of treatment adoption. The top panel of the figure plots the average point estimates (and standard deviations) from the standard event-study specification \eqref{eq:es}. Even though the data-generating process satisfies parallel trends, because the group$\times$period-specific average treatment effects are heterogeneous, the estimated leads of adoption exhibit a pre-treatment dip in outcomes, creating the mistaken appearance of a violation of parallel trends. This is consistent with the results of \citet{sun2020} and the discussion in Section \ref{sec:es}. The second panel summarizes estimates obtained by manually aggregating group$\times$duration-specific average treatment effects across groups, and the third summarizes estimates from the two-stage event-study procedure outlined in Section \ref{sec:es}. The aggregated and two-stage results are nearly identical; both present an accurate picture of pre-treatment trends, as well as the evolution of the effect of the treatment over its course (I present the means and standard deviations of the point estimates in Appendix Table \ref{tab:sim-es}).

\section{Empirical application}\label{sec:app}

To illustrate the application of the results presented above, I revisit Autor's \citeyearpar{autor2003} analysis of the effect of court rulings limiting the doctrine of employment at will on the growth of the temporary help services sector (THS). The key data are observations on the log of state-level THS employment and indicators for state-level legal exceptions to employment at will between 1979 and 1995. The baseline difference-in-differences specification regresses THS employment on an exception indicator and state and year effects. The estimate, reported in the top panel of Table \ref{tab:app-es}, of about .11 with a standard error (clustered on state) of about .1 replicates Autor's baseline result.\footnote{Autor's \citeyearpar{autor2003} preferred specifications include additional covariates and state-specific linear time trends, which I ignore for simplicity.}

\begin{figure}

\caption{Application DD weights\label{fig:app-wts}}

\medskip{}

\centering{}\includegraphics{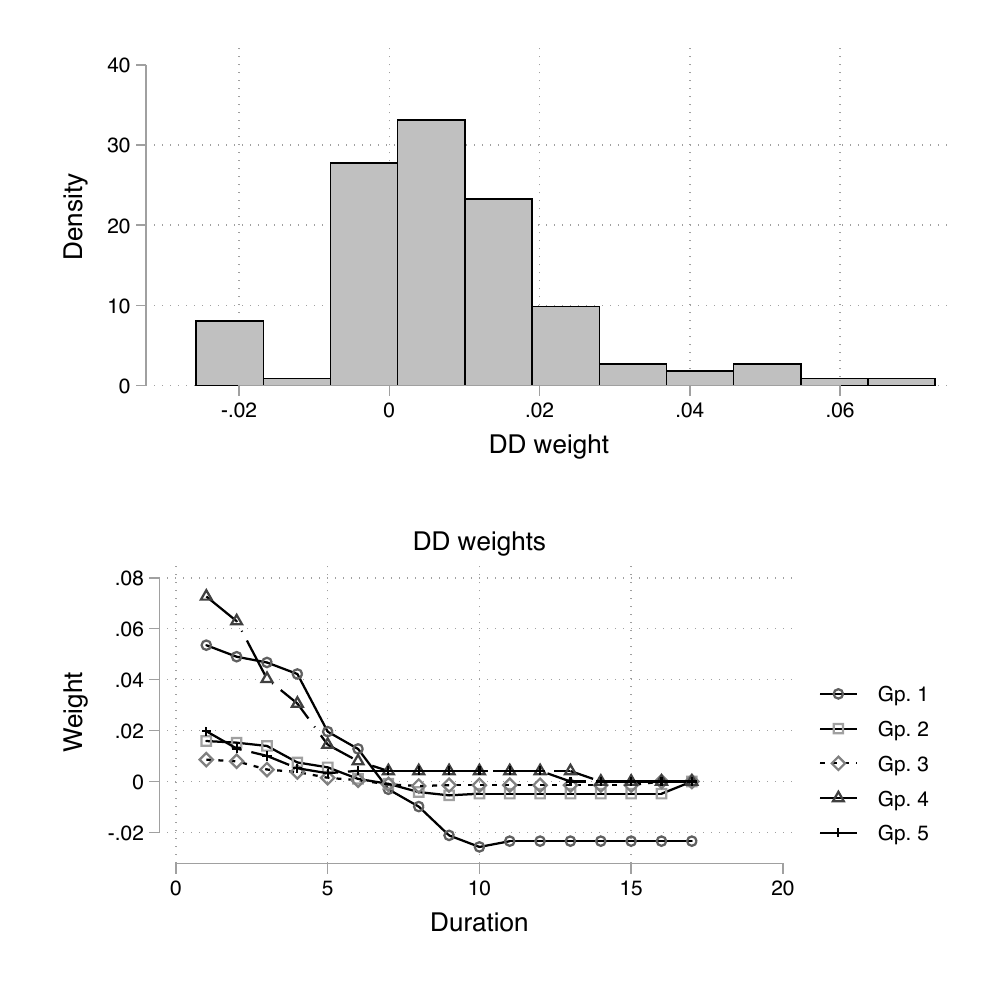}
\end{figure}

Following the discussion above, if the effect of the treatment is heterogeneous across treatment groups and periods, the difference-in-differences regression estimate represents a difficult-to-interpret weighted average of group$\times$period treatment effects. The first group in this application is treated in 1979, and the treatment is initiated for additional groups in every year through 1988, for a total of 10 groups. The difference-in-differences estimate therefore places nonzero weight on 125 group$\times$period-specific treatment effects. The top panel of Figure \ref{fig:app-wts} plots the distribution of these weights. Some are negative, although more are positive, and the positive magnitudes tend to exceed the negative ones. The distribution is also skewed, with a handful of effects receiving relatively large weights. The bottom panel of the figure plots the weights themselves for the first five groups. Consistent with the results in Section \ref{sec:dd-estimand} (and the literature cited there), differences in differences places more weight on treatment effects for groups that are treated earlier, that occur in earlier in the course of the treatment for each group, and for larger groups.

As a point of comparison to the difference-in-differences regression estimate, Table \ref{tab:app} also presents an estimate of the group$\times$period average treatment effect, obtained by estimating models with separate treatment indicators for each treated group and period, then averaging the coefficients on those indicators according to the empirical distribution of groups and periods among treated units.\footnote{For this, and all subsequent, estimates, I exclude units that were already treated in 1979, the first year of the sample, since no treatment effects are identified for this group. I calculate standard error for the aggregated estimator using the delta method, taking the distribution of groups and periods among the treated as fixed, and using state-clustered standard errors for the group$\times$period-specific regression coefficients.} The aggregated point estimate of about .09 is somewhat smaller than the difference-in-differences regression estimate. Since the results below (which agree with Autor's original findings) show that the effects of the treatment tend to be somewhat larger in the first few periods after the treatment, this is consistent with the weights illustrated above. At the same time, the broad similarity between the difference-in-difference and aggregated estimates suggests that the effects of the treatment are fairly homogeneous across groups and over time.

Table \ref{tab:app} also presents a two-stage difference-in-differences estimate of this treatment effect, obtained by estimating both equations simultaneously via GMM (and clustering standard errors at the state level). The two-stage estimate is similar to the aggregated estimate, illustrating the effectiveness of the two-stage approach. The table also presents the results from a version of the two-stage estimator in which a correctly specified first-stage equation that includes interactions between treatment status and period indicators is estimated using the full sample. The second-stage estimate is similar to the previous two-stage estimate (for which the first stage consists of a regression of untreated outcomes on state and time effects alone), although the standard error is slightly smaller.

As I note in Section \ref{sec:estimands}, the treatment effects estimated above are averaged across groups with different completed treatment durations. Because the shortest observed treatment duration is eight periods, it is possible to average the group-specific eight-period average treatment effects over all treatment groups. The aggregated estimate of this treatment effect is about .11. Consistent with the larger estimated early-duration treatment effects presented below, this is somewhat larger than the preceding estimate, which is averaged across all treated groups and periods. I also implement a stacked difference-in-differences estimator for this average effect, in which the dataset for each treatment group consists of observations on that group and the group of never-treated observations, one period before through eight periods after the adoption of the treatment. The resulting estimate is also about .11. Finally, I estimate this effect by the two-stage difference in differences by simply restricting the second-stage estimation sample to untreated observations and those with completed durations of eight periods or fewer. The two-stage and stacked estimates are also similar.

\begin{figure}
\caption{Application event studies\label{fig:app-es}}

\medskip{}

\begin{centering}
\includegraphics[height=0.9\textwidth]{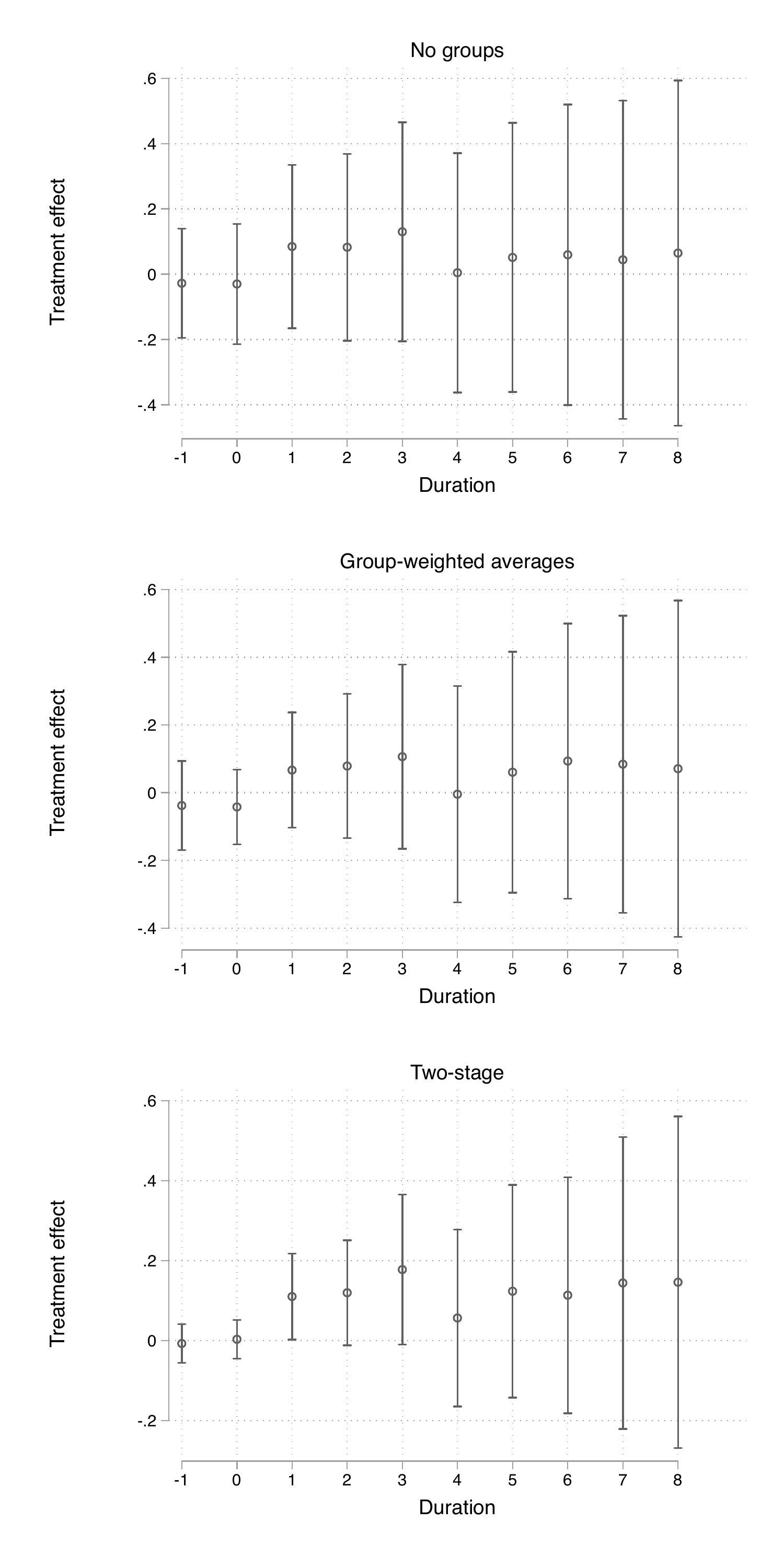}
\par\end{centering}
\medskip{}

Notes: Means and 95\% confidence intervals (based on state-clustered
standard errors).
\end{figure}

\cite{autor2003} also estimates an event-study regression to test the plausibility of parallel trends. To replicate this, I estimate a standard event-study regression using a version of specification \eqref{eq:es} that includes two leads of treatment adoption. The top panel of Figure \ref{fig:app-es} plots the coefficient estimates for these leads and the first eight post-treatment periods (I present the full set of point estimates in Appendix Table \ref{tab:app-es}). The estimates suggest that the effect of the treatment is larger in periods nearer adoption, stabilizing to a smaller value for durations of four or more periods. To examine the effect of heterogeneity on the event-study estimates, I also perform an aggregated event study using a model that includes separate duration-specific effects for each treatment group, following \citet{sun2020}.\footnote{To implement this, I exclude the second and third treatment groups, for which two leads of adoption are not observed, and average the first eight duration-specific effects over the remaining seven groups, weighted by cohort size.} The results, presented in the second panel of the figure, are similar to those from the event study that ignores treatment groups, suggesting that there is little heterogeneity in the duration-specific effects with respect to the timing of adoption. Finally, the bottom panel of the figure plots the results from a two-stage event study, obtained by replacing treatment status in the second stage of the procedure with indicators for each treatment duration. Once again, the two-stage estimates are similar to the aggregated estimates.

\section{Conclusion}\label{sec:conclusion}

When adoption of a treatment is staggered across time, and the average effects of the treatment vary by group and period, the usual difference-in-differences regression specification does not identify an easily interpretable measure of the typical effect of the treatment. When the duration-specific effects are also heterogeneous, neither do the coefficients from the usual event-study specification. The ultimate source of these identification failures is that outcomes are not necessarily linear in group, period and treatment status, as difference-in-differences and event-study regression specifications assume.

The two-stage approach developed in this paper is motivated by the observation that, under parallel trends, untreated outcomes are linear in group and period effects. Those effects are therefore identified from a first-stage regression estimated using the sample of untreated observations. The average effect of the treatment on the treated is then identified from a regression of outcomes on treatment status, after removing group and period effects. This procedure is robust to the presence of heterogeneous treatment effects when treatment adoption is staggered. It is also simple and intuitive, and can be extended to identify a variety of different treatment effect measures. Monte Carlo simulations and an empirical example show that the two-stage estimators correctly identify informative average treatment effect measures, in some cases outperforming alternative estimators that are also more difficult to implement.

\section*{Appendix A: Proofs}

\subsection*{Consistency and unbiasedness}

{\em Consistency.} The consistency of $(\hat{\lambda}_g, \hat{\gamma}_p)$ for $(\lambda_g, \gamma_p)$ follows from \eqref{eq:pt} and standard least-squares consistency arguments if $G$ and $P$ are fixed as the sample size grows. Otherwise, if the $Y_{gpit}$ represent sample averages calculated using $N_{gpit}$ observations, then under standard conditions, $Y_{gpit}\xrightarrow{p} E(Y_{gpit}|g,p,i,t)$ as $N_{gpit} \to \infty$. Since, under parallel trends, $E[E(Y_{gpit}|g,p,i,t)|g,p]=\lambda_g + \gamma_p$, $(\hat{\lambda}_g, \hat{\gamma}_p) \xrightarrow{p}(\lambda_g, \gamma_p)$ by the continuous mapping theorem. In either case, $Y_{gpit} - \hat{\lambda}_g - \hat{\lambda}_p \xrightarrow{p} E(Y_{gpit}|g,p,i,t) - \lambda_g - \gamma_p$. Since $E[E(Y_{gpit}|g,p,i,t)|g,p] - \lambda_g - \gamma_p = \beta_{gp} D_{gp}$ is linear in $D_{gp}$, the second stage coefficient is consistent for $E(\beta_{gp}|D_{gp}=1)$ by the continuous mapping theorem.	

{\em Unbiasedness.} From standard arguments applied to the first stage, $E[(\hat{\lambda}_g, \hat{\gamma}_t) | g, p, D_{gp}] = E[(\hat{\lambda}_g, \hat{\gamma}_t) | g, p] = (\lambda_g, \gamma_t)$. Hence we can write $Y_{gpit} - \hat{\lambda}_g - \hat{\gamma}_t = E(\beta_{gp} | D_{gp}=1) D_{gp} + U_{gpit}$ where $E(U_{gpit} | D_{gp})=E[E(U_{gpit}|g, p, D_{gp})|D_{gp}]=0$, which implies that the second-stage is unbiased for $E(\beta_{gp}|D_{gp}=1)$.

%{\em Unbiasedness.} From standard arguments, $E[(\hat{\lambda}_g, \hat{\gamma}_t) | g, p, D_{gp}=0] = E[(\hat{\lambda}_g, \hat{\gamma}_t) | g, p] = (\lambda_g, \gamma_t)$. Hence we can write $\tilde{Y}_{gpit} = E(\beta_{gp} | D_{gp}=1) D_{gp} + U_{gpit}$, where $E(U_{gpit}|g,p)=E[\tilde{Y}_{gpit} - E(\beta_{gp} | D_{gp}=1)D_{gp} | g, p]=0$. Then
%\begin{align*}
%E(\hat{\beta}^{2SDD} | g, p, D_{gp}) &= E\left( \frac{\sum_{i,t} D_{gpit}\tilde{Y}_{gpit}} 
%{\sum_{i,t} D_{gpit}^2} \bigg| g, p, D_{gp} \right) \\
%&= E(\beta_{gp} | D_{gp}=1) + \frac{\sum_{i,t} D_{gpit}E(U_{gpit}| g, p, D_{gp})}{\sum_{i,t} D_{gpit}^2} \\
%&= E(\beta_{gp} | D_{gp}=1).
%\end{align*}
%The proof the follows from iterating expectations.

\subsection*{The regression DD estimand \eqref{eq:omega}}

From \eqref{eq:pt}, we can write
\begin{equation}\label{eq:dgp}
  Y_{gpit} = \lambda_g + \gamma_p 
  + \sum_{h=1}^G \sum_{q=h}^{P} \beta_{hq} 1(h,q)_{gpit} 
  + \varepsilon_{gpit},
\end{equation}
where $1(h,q)_{gpit}$ is an indicator for whether observation $(g,p,i,t)$ corresponds to group $h$ and period $q$, and $E[\varepsilon_{gpit} | g, p, (1(h,q)_{gpit})] = 0$.

Let $\tilde{D}_{gp}$ denote the residual from a population regression of $D_{gp}$ on group and period fixed effects. By the Frisch-Waugh-Lovell theorem, the coefficient on $D_{gp}$ from a population regression of $Y_{gpit}$ on $D_{gp}$ and group and period effects is
\begin{align*}
  \beta^* &= \frac{E(\tilde{D}_{gp} Y_{gpit})}{E(\tilde{D}_{gp}^2)} \\
  &= \frac{E[\tilde{D}_{gp} \sum_{h=1}^G \sum_{q=h}^{P} \beta_{hq} 1(h,q)_{gpit}]}{E(\tilde{D}_{gp}^2)} \\
  &= \sum_{h=1}^G \sum_{q=h}^{P} \frac{E[\tilde{D}_{gp} 1(h,q)_{gpit}]}{E(\tilde{D}_{gp}^2)} \beta_{hq}  \\
  &= \sum_{g=1}^G \sum_{p=g}^{P} \omega_{gp} \beta_{gp}.
\end{align*}
where $\omega_{gp}$ is the coefficient from a regression of $1(h,q)_{gpit}$ on $D_{gp}$ and group and period fixed effects. The second equality uses the facts that $\varepsilon_{gpit}$ is mean-independent of the regressors and that $\tilde{D}_{gp}$ is uncorrelated with group and period effects by construction.\footnote{This, and the related result in \citet{sun2020}, can also be established by thinking of the term $\sum_{h=1}^G \sum_{q=h}^{P} \beta_{hq} 1(h,q)_{gpit}$ in \eqref{eq:dgp} as an omitted variable, and taking its projection onto the included regressors.}

The weight $\omega_{gp}$ that difference in differences places on $\beta_{gp}$ is the coefficient on $D_{gp}$ from a regression of $1(g,p)_{gpit}$ on $D_{gp}$ and group and period fixed effects. By the Frisch-Waugh-Lovell theorem, this is equivalent to the slope coefficient from a population regression of $1(g,p)_{gpit}$ on the residual from an auxiliary regression of $D_{gp}$ on group and period effects. Using the two-way within or double-demeaned transformation, this residual can be expressed as
\begin{equation}\label{eq:resid}
  \tilde{D}_{gp} = [D_{gp} - P(D_{gp}=1 | g)] - [P(D_{gp}=1|p)-P(D_{gp}=1)].
\end{equation}

Since $E(\tilde{D}_{gp}^2)=E(\tilde{D}_{gp} D_{gp})$, $\omega_{gp}$ can also be expressed as
\begin{align*}
  \omega_{gp} &= \frac{E(1(g,p)_{gpit}\tilde{D}_{gp})}{Var(\tilde{D}_{gp})} \\
    &= \frac{E(\tilde{D}_{gp} | 1(g,p)_{gpit}=1) P(1(g,p)_{gpit}=1)}
    {E(\tilde{D}_{gp} | D_{gp}=1) P(D_{gp}=1)} \\
    &= \frac{\{1-P(D_{gp}=1 | g)-[P(D_{gp}=1 | p) - P(D_{gp}=1)]\}P(g,p)}
    {\sum_{g=1}^G \sum_{p=g}^{P} \{1-P(D_{gp}=1 | g)-[P(D_{gp}=1 | p) - P(D_{gp}=1)]\}P(g,p)\}},
\end{align*}
where the final equality uses \eqref{eq:resid}.

\subsection{Stacked differences in differences}

In the stacked approach, a new dataset is created for each treated group, containing observations on that group $\bar{R}$ periods before, and $\bar{P}$ periods after, the treatment is adopted, as well as on units that are not yet treated during these periods. These group-specific datasets are stacked, and outcomes are regressed on treatment status and dataset-specific group and period fixed effects:
\[
  Y_{cgpit} = \lambda_{cg} + \lambda_{cp} + \beta D_{cgp} + \varepsilon_{cgpit},
\]
where $cgpit$ indexes the value of an observation in the dataset for group $c$ for the $i$th member of group $g$ during the $t$th time of period $p$.

Let $D_{cgp}$ be an indicator for whether group $g$ is treated during period $p$ of the group-$c$ dataset, and $D_{rcgp}$ be an indicator for whether members of $g$ have been treated for $r \in \{1, \dots, \bar{P}\}$ periods as of period $p$ in dataset $c$. Let $\tau = \bar{P}/(\bar{P}+\bar{R}+1)$ denote the fraction of periods during which treated units in any group-specific dataset are treated, $\pi_c$ denote the fraction of units in dataset $c$ that belong to the treatment group, and $\rho_c$ denote size of the group-$c$ dataset relative to the stacked dataset.

The weight $\omega_{rg}$ that stacked differences in differences places on the $r$-period average treatment effect $\beta_{rg}$ for group $g$ is given by the slope coefficient from a population regression of $D_{rcgp}$ on the residual $\tilde{D}_{cgp}$ from a regression of $D_{cgp}$ on dataset$\times$period and dataset$\times$group effects. This residual is
\[
  \tilde{D}_{cgp} = D_{cgp} - P(D_{cgp}=1 | g, c) - [P(D_{cgp}=1 | p, c) - P(D_{cgp}=1 | c)],
\]
where statements conditional on $c$ are true in the population corresponding to dataset $c$. Using this expression and adapting \eqref{eq:omega} to the stacked setting, 
\begin{align*}
  \omega_{rg} &= \frac{[1-\tau - (\pi_c - \tau \pi_c)] P(D_{rcgp}=1)}
   {\sum_{c=1}^G \sum_{p=1}^{\bar{P}} [1-\tau - (\pi_c - \tau \pi_c)] P(D_{rcgp}=1)} \\
   &= \frac{(1-\tau)(1-\pi_c) \tau \pi_c \rho_c}
   {\sum_{c=1}^G \sum_{p=1}^{\bar{P}} (1-\tau)(1-\pi_c) \tau \pi_c \rho_c} \\
   &= \frac{(1-\pi_c) \pi_c \rho_c}
   {\bar{P} \sum_{c=1}^G (1-\pi_c) \pi_c \rho_c}.
\end{align*}

\section*{Appendix B: Stata syntax}

Suppose that \verb|y| refers to the outcome, \verb|year| the year, \verb|id| the group, and \verb|d| treatment status. The two-stage difference-in-differences estimator can be obtained, along with valid cluster-robust asymptotic standard errors, via GMM using the single Stata command:

\begin{verbatim}
gmm (eq1: (y - {xb: i.year} - {xg: ibn.id})*(1-d)) ///
  (eq2: y - {xb:} - {xg:} - {delta}*d), ///
  instruments(eq1: i.year ibn.id) ///
  instruments(eq2: d) winitial(identity) ///
  onestep quickderivatives vce(cluster id)	
\end{verbatim}
Variations on the two-stage estimator (such as the the two-stage event-study estimator) can be obtained using similar syntax.

\begin{table}

\caption{Simulation results\label{tab:sim}}

\medskip{}

\begin{centering}
\begin{tabular}{llccc}
\hline 
 &  & Simulation 1 &  & Simulation 2\tabularnewline
\cline{3-3} \cline{5-5} 
All periods & True & 4.08 &  & 3.46\tabularnewline
 & Diff-in-diff & 3.51 &  & 2.72\tabularnewline
 &  & (0.28) &  & (0.23)\tabularnewline
 & Aggregated & 4.12 &  & 3.48\tabularnewline
 &  & (0.28) &  & (0.22)\tabularnewline
 & Two-stage & 4.12 &  & 3.48\tabularnewline
 &  & (0.28) &  & (0.22)\tabularnewline
 &  &  &  & \tabularnewline
Four-period & True & 3.17 &  & 2.75\tabularnewline
 & Aggregated & 3.21 &  & 2.78\tabularnewline
 &  & (0.32) &  & (0.25)\tabularnewline
 & Stacked & 3.21 &  & 2.87\tabularnewline
 &  & (0.33) &  & (0.26)\tabularnewline
 & Two-stage & 3.21 &  & 2.78\tabularnewline
 &  & (0.32) &  & (0.25)\tabularnewline
\hline 
\end{tabular}
\par\end{centering}
\medskip{}

Notes: Means and standard deviations from 250 simulations.

\end{table}

\clearpage{}

\clearpage{}

\clearpage{}

\clearpage{}

\begin{table}
\caption{Application estimates\label{tab:app}}

\medskip{}

\begin{centering}
\begin{tabular}{llc}
\hline 
All periods & Diff-in-diff & 0.112\tabularnewline
 &  & (0.096)\tabularnewline
 & Aggregated & 0.087\tabularnewline
 &  & (0.158)\tabularnewline
 & Two-stage & 0.092\tabularnewline
 &  & (0.157)\tabularnewline
 & Two-stage (incl. treat$\times$time & 0.097\tabularnewline
 & $\quad$in first stage) & (0.149)\tabularnewline
 &  & \tabularnewline
Eight periods & Aggregated & 0.106\tabularnewline
 &  & (0.126)\tabularnewline
 & Stacked & 0.114\tabularnewline
 &  & (0.121)\tabularnewline
 & Two-stage & 0.110\tabularnewline
 &  & (0.126)\tabularnewline
\hline 
\end{tabular}
\par\end{centering}
\medskip{}

Notes: Standard errors clustered on state. Standard errors for aggregated
estimators calculated using the delta method, assuming that the treated
distribution of groups and periods is fixed. Two-stage estimates computed
by estimating both equations simultaneously by GMM.
\end{table}

\clearpage{}

\clearpage{}

\section*{Appendix C: Event-study estimates}

\begin{table}[h]
\caption{Simulation event-study results\label{tab:sim-es}}

\medskip{}

\begin{centering}
\begin{tabular}{cccccccc}
\hline 
 & \multicolumn{3}{c}{Simulation 1} &  & \multicolumn{3}{c}{Simulation 2}\tabularnewline
\cline{2-4} \cline{3-4} \cline{4-4} \cline{6-8} \cline{7-8} \cline{8-8} 
Period & Aggregated & No groups & Two-stage &  & Aggregated & No groups & Two-stage\tabularnewline
\cline{1-4} \cline{2-4} \cline{3-4} \cline{4-4} \cline{6-8} \cline{7-8} \cline{8-8} 
-1 & 0.024 & -0.425 & 0.018 &  & 0.003 & -0.396 & 0.007\tabularnewline
 & (0.385) & (0.372) & (0.239) &  & (0.248) & (0.236) & (0.147)\tabularnewline
0 & 0.000 & -0.540 & -0.006 &  & -0.003 & -0.607 & -0.004\tabularnewline
 & (0.360) & (0.357) & (0.212) &  & (0.288) & (0.286) & (0.138)\tabularnewline
1 & 1.246 & 0.645 & 1.239 &  & 1.019 & 0.317 & 1.016\tabularnewline
 & (0.616) & (0.619) & (0.590) &  & (0.464) & (0.458) & (0.439)\tabularnewline
2 & 2.295 & 1.706 & 2.286 &  & 2.038 & 1.401 & 2.036\tabularnewline
 & (0.626) & (0.623) & (0.592) &  & (0.460) & (0.466) & (0.441)\tabularnewline
3 & 4.075 & 3.526 & 4.068 &  & 3.527 & 2.986 & 3.527\tabularnewline
 & (0.607) & (0.605) & (0.581) &  & (0.460) & (0.481) & (0.439)\tabularnewline
4 & 5.255 & 4.777 & 5.249 &  & 4.531 & 4.091 & 4.531\tabularnewline
 & (0.700) & (0.695) & (0.666) &  & (0.478) & (0.471) & (0.462)\tabularnewline
\hline 
\end{tabular}
\par\end{centering}
\medskip{}

Notes: Means and standard deviations from 250 simulations.
\end{table}

\clearpage{}

\begin{table}
\caption{Application event study estimates\label{tab:app-es}}

\medskip{}

\begin{centering}
\begin{tabular}{cccc}
\hline 
Period & Aggregated & No groups & Two-stage\tabularnewline
\hline 
-1 & -0.042 & -0.030 & 0.003\tabularnewline
 & (0.055) & (0.092) & (0.024)\tabularnewline
0 & -0.038 & -0.028 & -0.007\tabularnewline
 & (0.066) & (0.084) & (0.024)\tabularnewline
1 & 0.067 & 0.084 & 0.110\tabularnewline
 & (0.085) & (0.125) & (0.054)\tabularnewline
2 & 0.078 & 0.082 & 0.119\tabularnewline
 & (0.106) & (0.143) & (0.066)\tabularnewline
3 & 0.106 & 0.130 & 0.177\tabularnewline
 & (0.136) & (0.168) & (0.094)\tabularnewline
4 & -0.005 & 0.004 & 0.056\tabularnewline
 & (0.160) & (0.183) & (0.111)\tabularnewline
5 & 0.060 & 0.051 & 0.123\tabularnewline
 & (0.178) & (0.206) & (0.133)\tabularnewline
6 & 0.093 & 0.059 & 0.113\tabularnewline
 & (0.203) & (0.230) & (0.148)\tabularnewline
7 & 0.084 & 0.044 & 0.144\tabularnewline
 & (0.219) & (0.244) & (0.182)\tabularnewline
8 & 0.071 & 0.065 & 0.146\tabularnewline
 & (0.248) & (0.244) & (0.207)\tabularnewline
\hline 
\end{tabular}
\par\end{centering}
\medskip{}

Notes: Standard errors clustered on state. Standard errors for aggregated
estimators calculated using the delta method, assuming that the treated
distribution of groups and periods is fixed. Two-stage estimates computed
by estimating both equations simultaneously by GMM.
\end{table}

%%%%%%%%%%%%%%%%%%%%%%%%%%%%%%%%%%%%%%%%%%%%%%%%%%%%%%%%%%%%%%%%%%%%%%%%%%%%%%%%
\end{document}